\documentclass[draft,jgrga]{AGUTeX}
\usepackage{graphicx}
\usepackage{amsfonts}
\usepackage{color}
\authorrunninghead{XIONG ET AL.}
\titlerunninghead{MC-MC INTERACTION AND ITS GEOEFFECTIVENESS: 2.}

\authoraddr{Ming Xiong, Huinan Zheng, and Shui Wang,
School of Earth and Space Sciences, University of Science and
Technology of China, Hefei, Anhui 230026, China.
(mxiong@ustc.edu.cn; hue@ustc.edu.cn; and swan@ustc.edu.cn)}

\begin{document}
\setkeys{Gin}{draft=false} %Please uncomment it for official format

\title{Magnetohydrodynamic simulation of the
interaction between two interplanetary magnetic clouds and its
consequent geoeffectiveness: 2. Oblique collision}

\author{Ming Xiong, \altaffilmark{1} Huinan Zheng, \altaffilmark{1} and Shui Wang\altaffilmark{1}}

\altaffiltext{1}
{Chinese Academy of Sciences Key Laboratory for Basic Plasma Physics, School of Earth
and Space Sciences, University of Science and Technology of China, Hefei, Anhui 230026, China}

\begin{abstract}
The numerical studies of the interplanetary coupling between
multiple magnetic clouds (MCs) are continued by a 2.5-dimensional
ideal magnetohydrodynamic (MHD) model in the heliospheric
meridional plane. The interplanetary direct collision (DC) /
oblique collision (OC) between both MCs results from their
same/different initial propagation orientations. Here the OC is
explored in contrast to the results of the DC (Xiong et al.,
2007). Both the slow MC1 and fast MC2 are consequently injected
from the different heliospheric latitudes to form a compound
stream during the interplanetary propagation. The MC1 and MC2
undergo contrary deflections during the process of oblique
collision. Their deflection angles of $|\delta \theta_1|$ and
$|\delta \theta_2|$ continuously increase until both MC-driven
shock fronts are merged into a stronger compound one. The $|\delta
\theta_1|$, $|\delta \theta_2|$, and total deflection angle
$\Delta \theta$ ($\Delta \theta = |\delta \theta_1| + |\delta
\theta_2|$) reach their corresponding maxima when the initial
eruptions of both MCs are at an appropriate angular difference.
Moreover, with the increase of MC2's initial speed, the OC becomes
more intense, and the enhancement of $\delta \theta_1$ is much
more sensitive to $\delta \theta_2$. The $|\delta\theta_1|$ is
generally far less than the $|\delta\theta_2|$, and the unusual
case of $|\delta\theta_1|\simeq|\delta\theta_2|$ only occurs for
an extremely violent OC. But because of the elasticity of the MC
body to buffer the collision, this deflection would gradually
approach an asymptotic degree. As a result, the opposite
deflection between the two MCs, together with the inherent
magnetic elasticity of each MC, could efficiently relieve the
external compression for the OC in the interplanetary space. Such
deflection effect for the OC case is essentially absent for the DC
case. Therefore, besides the magnetic elasticity, magnetic
helicity, and reciprocal compression, the deflection due to the OC
should be considered for the evolution and ensuing
geoeffectiveness of interplanetary interaction among successive
coronal mass ejections (CMEs).
\end{abstract}
\begin{article}

%% ------------------------------------------------------------------------ %%
%% ------------------------------------------------------------------------ %%

\section{Introduction}
One of the greatest concerns within the current space science
community has been increasingly focused on the Sun-Earth system,
which is intimately linked by the solar wind. The solar wind
originates from the chromospheric network
\citep{Xia2003a,Xia2003b,Xia2004}, according to the measurements
of ultraviolet emission and Doppler shifting speed in the inner
corona, carries non-Wentzel-Kramers-Brillouin Alfv\'en Waves in
the differential flow of multiple ion species
\citep{Li2007a,Li2008}, is very likely driven by an ion cyclotron
resonance mechanism via the Kolmogorov turbulent cascade
\citep{Li2004}, and transports the angular momentum from the Sun
\citep{Li2006,Li2007b,Li2009}. The ubiquitous interplanetary solar
wind highly fluctuates, owing to outward-emanating disturbances
from solar activities. Therefore, the Sun serves as the driver for
the cause-and-effect transporting chain of space weather.

The interplanetary space, which \citet{Dryer1994} calls a
``transmission channel" between the Sun and the Earth , is a
nonlinear system consisting of various discontinuous fronts,
diffusion processes, and couplings between different spatial and
temporal scales. A magnetic flux rope levitating in the corona may
suddenly lose its equilibrium and consequently escape into the
interplanetary space \citep{Chen2006,Chen2007}. The interplanetary
manifestation of such a magnetic rope is identified as a magnetic
cloud (MC) with enhanced magnetic field magnitude, smooth rotation
of the magnetic field vector, and low proton temperature
\citep{Burlaga1981}. The passage of an MC across the Earth
triggers a significant geomagnetic storm because of large
southward magnetic flux in the MC body
\citep{Tsurutani1988,Gosling1991}. Hence MCs are an important
subset of interplanetary CMEs (ICMEs), whose fraction is $\sim$
100\%, though with low statistics, at solar minimum and $\sim$
15\% at solar maximum \citep{Richardson2004}. Especially at solar
maximum, when the daily occurrence rate of CMEs is about 4.3 on
average based on the {\it SOHO/Lasco} CME catalogue
(http://cdaw.gsfc.nasa.gov/CME\_list), CMEs very likely interact
with each other on their journey toward the Earth. Two distinct
observation events of interaction between an early slow CME1 and a
late fast CME2 within 30 solar radii were presented by
\citet{Gopalswamy2002b}: (1) two fast CMEs on 4 November 1997,
which were initially $100^\circ$ apart in relation to their source
regions according to {\it Yohkoh/SXT} observation, led to the
plowing of the CME2-driven shock through the CME1 in the
field-of-view (FOV) of {\it Lasco C2/C3}; (2) two fast CMEs from
the same source region on 20 January 2001, initially two hours
apart, were later indistinguishable in the FOV of {\it Lasco C3},
and are therefore thought to have cannibalized each other. As the
coupling of multiple CMEs from the same/different heliographic
location of source region is defined as the direct collision (DC)
/ oblique collision (OC) by \citet{Xiong2006b}, these two events
of 4 November 1997 and 20 January 2001 are the cases of OC and DC,
respectively. The radio signatures of coronal mass ejection
cannibalism typically precede the intersection of the leading-edge
trajectories and behave as an intense continuum-like radio
emission enhancement, usually following a type II radio burst on
basis of {\it Wind/WAVES} observation
\citep{Gopalswamy2001,Gopalswamy2002a}. \citet{Gopalswamy2002a}
argue that the nonthermal electrons responsible for this new type
of ratio emission are accelerated due to magnetic reconnection
between two CMEs and/or the formation of a new shock at the time
of collision between two CMEs. Meanwhile, some interplanetary
complicated structures were also reported in the near-Earth space,
such as the complex ejecta \citep{Burlaga2002}, compound stream
\citep{Burlaga1987,Wang2003a,Dasso2009}, shock-penetrated MCs
\citep{Lepping1997,Wang2003b,Berdichevsky2005}, and
non-pressure-balanced ``MC boundary layers'' associated with
magnetic reconnection \citep{Wei2003,Wei2006}. According to the in
situ observations of spacecraft at 1 AU, the evolutionary
signatures of ICMEs' interaction include heating of the plasma,
acceleration/deceleration of the leading/trailing ejecta,
compressed field and plasma in the leading ejecta, possible
disappearance of shocks, and strengthening of the shock driven by
the accelerated ejecta \citep{Farrugia2004}. Since magnetic
diffusion in interplanetary space is much less than that in the
solar corona, the cannibalism of CMEs that interact in the {\it
Lasco} FOV \citep{Gopalswamy2001,Gopalswamy2002a} should not occur
in the interplanetary space \citep{Xiong2007a}. Moreover, formed
by multiple CMEs/ICMEs colliding, the compound stream at 1 AU
could be in a different evolutionary stage. The position of the
overtaking shock at 1 AU can be (1) still in the MC, such as an 18
October 1995 event \citep{Lepping1997} and a 5-7 November 2001
event \citep{Wang2003b}, or (2) ahead of the MC after ultimately
penetrating it \citep{Berdichevsky2005}. The compressed magnetic
field downstream of the shock front is northward for the 18
October 1995 event \citep{Lepping1997} and southward for the 5-7
November 2001 event \citep{Wang2003b}. Therefore, the latter event
of 5-7 November 2001 resulted in a great magnetic storm of
$Dst\approx-300$ nT. An important interplanetary origin for the
great geomagnetic storms have already been identified by the
observations \citep{Wang2003a,Farrugia2006,Dasso2009} and
simulations \citep{Xiong2006a,Xiong2006b,Xiong2007a,Xiong2007b} as
multi-ICME structures, accompanying intense compression of
southward magnetic flux during the interaction process. When the
compound structure reaches the Earth through the interplanetary
space, its physical parameters are jointly decided by three
factors: (1) individual CMEs themselves, (2) inhomogeneous
interplanetary medium, (3) irreversible interacting process among
these CMEs/ICMEs \citep{Xiong2007b}. Due to the intractability of
analytical reduction, compound structures resulting from the
interaction of multiple CMEs/ICMEs have been extensively studied
in numerical simulations: e.g., complex ejecta \citep{Xiong2005},
interaction of a shock wave with an MC
\citep{Vandas1997,Xiong2006a,Xiong2006b}, and coupling of multiple
MCs \citep{Schmidt2004,Lugaz2005,Xiong2007a}. Particularly,
\citet{Xiong2005,Xiong2006a,Xiong2006b,Xiong2007a} and
\citet{Xiong2007b} conducted a systematic and delicate numerical
MHD simulation of interplanetary compound structures in terms of
their formation, propagation, evolution, and ensuing
geoeffectiveness. These simulation works do well provide
theoretical interpretations for physical phenomena of compound
structures observed by the {\it SOHO}, {\it Wind}, and {\it ACE}
spacecraft.

The radial lift-off of a CME at its onset phase from a solar
source region sometimes deviates from the radial ray during its
outward movement. The non-straight trajectory substantiates that
deflections do happen during CME/ICME propagation. The deflection
effect plays a notable role in space weather predicting, since the
first step of prediction is whether or not a solar eruption will
ultimately affect the geospace environment \citep{Williamson2000}.
The near-Sun trajectory of a CME can be directly imaged by remote
sensing of a white light coronagraph onboard such spacecraft as
{\it Skylab}, {\it SOHO}, and {\it STEREO}. \citet{MacQueen1986}
found that 29 CME events observed during the Skylab epoch of solar
minimum from 1973 to 1974 underwent an average $2.2^\circ$
equatorward deflection, and ascribed that the deflection to the
nonradial forces arising from the background coronal magnetic and
flow patterns. \citet{Cremades2004} identified the CME events from
{\it SOHO/Lasco} FOV and their corresponding source regions from
the {\it SOHO/EIT} and {\it SOHO/MDI} from January 1996 to
December 2002, and found that the position angle (PA) of {\it
Lasco}-imaged CMEs deviates statistically about $18.6^\circ$
southward toward the lower latitude at solar minimum. They also
ascribed such equatorward deflection of CMEs from solar activity
belts to the surrounding fast solar wind from polar coronal holes
with a stronger total plasma and magnetic field pressure.
\citet{Gopalswamy2001} reported that on 10 June 2000, a slow CME
of 290 km/s was overtaken by a fast CME of 660 km/s from a
different solar source region; the core of the slow CME was
leftward deviated by $13^\circ$ in terms of the PA in the {\it
Lasco/C3} FOV. \citet{Zhang2004} also reported a nonradial motion
of a gradually accelerated CME on 19 October 1997 from the $\it
SOHO$ observation. This peculiar CME was initiated above the east
limb at northern latitude $14^\circ$N in the {\it EIT} FOV, tilted
towards the equator as it rose in the {\it Lasco/C1} FOV, and was
very symmetric with respect to the equator later in the {\it Lasco
C2/C3} FOV. Furthermore, besides the occurrence within the {\it
Lasco/C3} FOV, the CME/ICME deflection does exist beyond the
near-Sun space. On the basis of statistical analyses of
interplanetary scintillation observations, \citet{Wei1988} and
\citet{Wei1991} found that the solar-flare-generated shock
deflects eastward in the heliospheric equator and equatorward in
the heliospheric meridian during its interplanetary propagation.
This deflection evidence of interplanetary shock aphelion results
from joint effects of the (1) spiral interplanetary magnetic field
(IMF), (2) westward movement of the heliographic location of a
solar flare during the impulsive phase, and (3) heterogenous
medium consisting of the fast solar wind from open corona magnetic
field and the slow solar wind astride the heliospheric current
sheet \citep{Hu1998,Hu2001}. In addition, the solar source
distribution of Earth-encountered halo CMEs is east-west asymmetry
\citep{Wang2002,Zhang2003}. Some eastern limb CMEs hit the Earth
\citep{Zhang2003}, and conversely some disk CMEs missed the Earth
\citep{Schwenn2005}. According to an ICME's kinematic model
\citep{Wang2004}, ICMEs could be deflected as much as several tens
of degrees during its propagation by the background solar wind and
spiral IMF; a fast CME will be blocked by the background solar
wind ahead and deflected to the east; a slow CME will be pushed by
the following background solar wind and deviated to the west. The
existence of ICME deflection is obviously implied from the
evidence of indirect observations about the correlations between
the near-Sun CME and near-Earth ICME. However, direct observations
covering the entire interplanetary space have only been available
since the launching of {\it SMEI} and {\it STEREO} in the
twenty-first century. Most of the current spaceborne observations
are still heavily concentrated to the thirty solar radii by remote
sensing, and the geospace by in situ detecting. As interplanetary
observation data is relatively small, numerical simulations are
necessary and significant for understanding the whole of
interplanetary dynamics, including the deflection effect.
\citet{Xiong2006b} proposed that (1) the OC between a preceding MC
and a following shock results in the simultaneous opposite
deflections of the MC body and shock aphelion; (2) an appropriate
angular difference between the initial eruption of an MC and an
overtaking shock leads to the maximum deflection of the MC body;
(3) the larger the shock intensity is, the greater the deflection
angle. As a straightforward analogy to the MC-shock OC
\citep{Xiong2006b}, the interplanetary deflection can be also
expected for the MC-MC OC. As a result of collision of one MC with
either a shock or another MC, the deflections can be ascribed to
the interaction between different interplanetary disturbances. In
contrast to our models, the previous deflection models
\citep[e.g.,][]{Hu1998,Hu2001,Wang2004} are caused by the
interaction between the ambient solar wind and interplanetary
disturbance.

The conjecture about the interplanetary deflection from the MC-MC OC
is investigated in this paper. In addition, a simplified
circumstance of MC-MC DC, excluding the deflection effect, has
already been studied by \citet{Xiong2007a}. The following
conclusions are revealed from the MC-MC DC \citep{Xiong2007a}: (1)
when the accumulated magnetic elasticity can balance the external
colliding, the compressibility of double MCs reaches its maximum;
(2) this cutoff limit of compressibility mainly decides the
maximally available geoeffectiveness of double MCs, because
geoeffectiveness enhancement of MCs' interacting is ascribed to
compression; (3) the magnetic elasticity, magnetic helicity of each
MC, and compression between each are the key physical factors for
the formation, propagation, evolution, and resulting
geoeffectiveness of interplanetary double MCs. Here the study of
MC-MC OC is a more reasonable extension of that of MC-MC DC
\citep{Xiong2007a}. Thus two issues are naturally raised: (1) What
is the difference between the MC-MC DC and MC-MC OC in terms of the
interplanetary dynamics and ensuing geoeffectiveness? (2) Does such
a deflection effect caused by the MC-MC OC play a significant or
negligible role during interaction process? The answers to these
questions are explored by a 2.5-dimensional (2.5-D) numerical model
in ideal MHD process.

The present work targets the OC between two MCs as our logical
continuation in a series of studies for the interplanetary
compound structures
\citep{Xiong2005,Xiong2006a,Xiong2006b,Xiong2007a,Xiong2007b}. We
give the numerical MHD model in section \ref{Sec:Method}, describe
the dynamics and geoeffectiveness of two typical cases of double
MCs in section \ref{Sec:MC-MC}, analyze the roles of eruption
interval in section \ref{Sec:EID}, angular difference in section
\ref{Sec:ADD}, and collision intensity in section \ref{Sec:CID}
for two MCs' interacting, and summarize the paper in section
\ref{Sec:Conclusion}.

%% ------------------------------------------------------------------------ %%
%
%  SECTION HEADS
%
%% ------------------------------------------------------------------------ %%
\section{Numerical MHD Model}\label{Sec:Method}
The dynamics and geoeffectiveness of interplanetary compound
structures have already been numerically investigated by our
effective numerical model
\citep{Xiong2006a,Xiong2006b,Xiong2007a,Xiong2007b}. This model
quantitatively relates the output of solar disturbances at 25
$R_s$ to the interplanetary parameters and geomagnetic storm at 1
AU, thus establishing a cause-and-effect transporting chain for a
solar-terrestrial physical process. The concrete implementation of
this numerical model consists of two steps: (1) the numerical MHD
simulation of interplanetary disturbance propagation, and (2)
using the Burton empirical formula for the solar wind -
magnetosphere - ionosphere coupling to evaluate the geomagnetic
storm index $Dst$ \citep{Burton1975}. The detailed description of
the numerical model, including the numerical algorithm,
computational grid layout, ambient solar wind, is given in
\citet{Xiong2006a}.

An incidental MC, radially launched from the solar surface, is
characterized by several parameters: the emergence speed $v_{mc}$,
latitude $\theta_{mc}$, and time $t_{mc}$, et al. The following
MC2's emergence latitude $\theta_{mc2}$ is included for parametric
study in contrast to the DC along the heliospheric equator
\citep{Xiong2007a}. Both the MCs are consequently injected into
the simulation domain through a particular modification of the
inner boundary condition at 25 $R_s$
\citep{Vandas1995,Xiong2006a}. The DC and OC in the interplanetary
medium correspond to $\theta_{mc2} = 0^\circ$ and $\theta_{mc2}
\neq 0^\circ$, since the preceding MC1 emerges from the equator,
$\theta_{mc1} = 0^\circ$. Moreover, the MC2-driven shock in all of
our simulation cases is faster than the local magnetosonic speed
at any time in order to prevent weak shock dissipation in the MC
body of low plasma $\beta$.

\section{MC1-MC2 Interaction}\label{Sec:MC-MC}
All thirty-four cases of double MCs' interacting are assembled
into three groups in Table \ref{Tab2}, with ten cases of an
individual MC in two groups from Table \ref{Tab1} for comparison.
In Table \ref{Tab2}, the helicity of one MC $H_{mc1}=1$ is
opposite to that of the other MC $H_{mc2}=-1$, because we are only
interested in the maximally available geoeffectiveness among all
combinations of each MC helicity \citep{Xiong2007a}. Groups of an
individual-preceding MC (IPM), an individual-following MC (IFM),
an eruption-interval dependence (EID), an angular-difference
dependence (ADD), and a collision-intensity dependence (CID) are
studied, with case E$_2$ shared by Groups EID, ADD, and CID. The
slow MC1 of $v_{mc1}=400$ km/s, $H_{mc1}=1$,
$\theta_{mc1}=0^\circ$, and $t_{mc1}=0$ hour is chased and pounded
by a fast MC2 of various parameters. Here the parametric studies
of double MCs cover a wide spectrum of $t_{mc2}=10.2 \sim 44.1$
hours in group EID, $\theta_{mc2}=0^\circ \sim 50^\circ$ in group
ADD, and $v_{mc2}=450 \sim 1200$ km/s in group CID. Moreover, by
adjusting $Dt$ ($Dt=t_{mc1}-t_{mc2}$, $t_{mc1}=0$ hour), the
initiation delay between the two MC emergences in group EID, an
interplanetary compound stream consisting of double MCs may reach
a different evolutionary stage when it arrives at 1 AU. The
$t_{mc2}$ is prescribed to be 12.2 hours in groups ADD and CID for
the full development of double MCs' interacting within 1 AU. In
the following we address case E$_1$ of 30.1 hours and case E$_2$
of 12.2 hours in group EID, which are typical examples of double
MCs in the early and late evolutionary stages.

\subsection{Case E$_1$}\label{Sec:Case1}
\begin{figure*}
\noindent
  \includegraphics[width=0.93\textheight,angle=90]{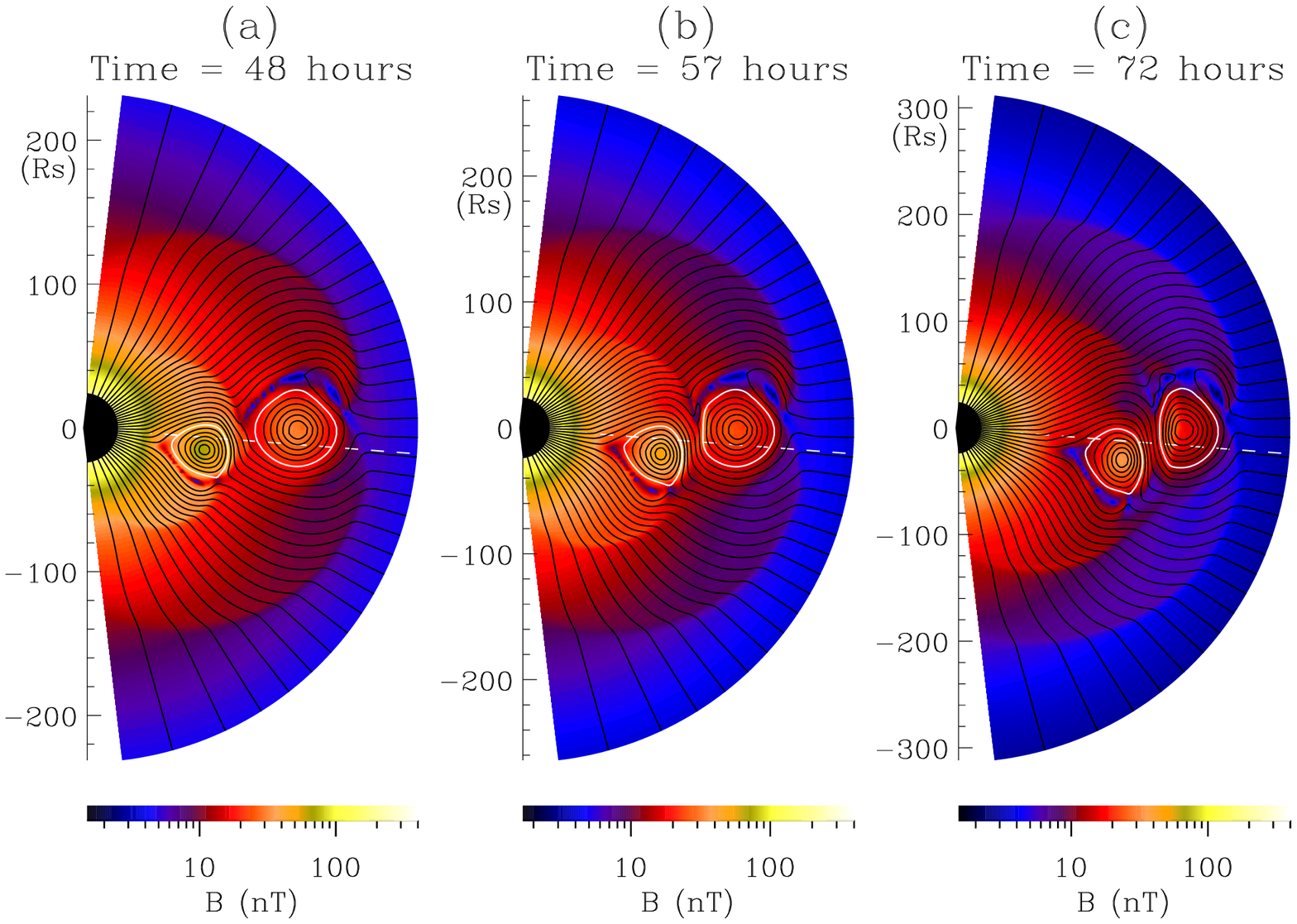}
\caption{} \label{case1-B}
%\end{landscapefigure*}
\end{figure*}

Figures \ref{case1-B}-\ref{case1-Line} shows the consequent behavior
of MC1-MC2 interaction of Case E$_1$ with the eruption speed
$v_{mc1}=$ 400 km/s, $v_{mc2}=$ 600 km/s, and the initiation delay
$t_{mc2}=$ 30.1 hours. The magnetic field lines, of which two are
enclosed with white solid lines marking the boundaries of MC1 and
MC2, are superimposed on each image of Figures
\ref{case1-B}-\ref{case1-Cf}. Two radial profiles, one through the
equator (noted by Lat. = $0^\circ$), the other through $4.5^\circ$
southward (white dashed lines in Figures
\ref{case1-B}-\ref{case1-Cf}, noted by Lat. = $4.5^\circ$S), are
plotted in Figure \ref{case1-Line}. The magnitude $B$ of the
magnetic field in the radial profile of Figures \ref{case1-B}a-c is
presented by subtracting its initial value $B|_{t=0}$ of ambient
equilibrium. The coupling of two MCs could be considered a
comprehensive interaction between two systems, each comprised of an
MC body and its driven shock. The MC2-driven shock and MC2 body are
successively involved in the interaction with the MC1 body. The
MC2-driven shock catches up with the MC1 body tail at 48 hours, as
seen in Figures \ref{case1-Vr}d and \ref{case1-Line}d. Across the
shock front, impending collision is influenced by the abrupt jump of
radial speed $v_r$ from 430 to 650 km/s. From then on, both MCs are
coupled with each other to form an interplanetary compound stream of
double MCs \citep{Wang2003a,Dasso2009}. At 57 hours, the marching
MC2-driven shock front behaves as a steep speed jump at MC1's rear
part (Figures \ref{case1-Vr}e and \ref{case1-Line}e), just
downstream from which the magnetic magnitude $B$ (Figures
\ref{case1-B}b and \ref{case1-Line}b) and fast magnetosonic mode
speed $c_f$ (Figures \ref{case1-Cf}h and \ref{case1-Line}h) are
locally enhanced. Due to the large initial delay $t_{mc2} = 30.1$
hours, only the rear half of the MC1 body is swept and compressed by
the MC2-driven shock within 1 AU (Figures \ref{case1-Vr}f and
\ref{case1-Line}f).

The in situ observation along Lat. $=4.5^\circ$S by a hypothetical
spacecraft at the Lagrangian point (L1) is shown in Figure
\ref{Satellite-1}. The boundary and core of each MC are identified
as dashed and solid lines, respectively. The rear half of the MC1
body is significantly gripped by the penetration of the MC2-driven
shock at the MC1 core and the push of the MC2 body upon the MC1
tail. The duration of MC1's rear half (9 hours) is much less than
that of MC1's anterior half (16 hours). The dawn-dusk electric
field $VB_z$ is swiftly intensified from 0 at 73 hours to $-13$
mV/m at 76 hours. Because the orientation of the magnetic field
within the double flux-rope structure is north-south-south-north,
the superposition of three individual southward $B_s$ regions from
the MC1, IMF, and MC2 behaves as a long-lived geoeffective solar
wind flow from 73 to 93 hours (Figure \ref{Satellite-1}d), and
results in a one-dip curve of $Dst$ with its minimum $-234$ nT at
87 hours (Figure \ref{Satellite-1}e).

\subsection{Case E$_2$}\label{Sec:Case2}
In case E$_2$, a much earlier emergence time of the MC2 ($t_{mc2}=$
12.2 hours) guarantees the full interaction between the two MCs
before their arrival at 1 AU. Only the evolution of $v_r$ is given
in Figures \ref{case2-Vr} and \ref{case2-Line} to visualize the
structure of double MCs. The initial emergence latitudes of MC1
($\theta_{mc1}|_{t=0}$) and MC2 ($\theta_{mc2}|_{t=0}$) are two
important parameters of solar eruption output. The nonzero
difference $D\theta|_{t=0}$
($D\theta|_{t=0}=\theta_{mc2}|_{t=0}-\theta_{mc1}|_{t=0}$) decides a
consequent OC in the interplanetary space. The collision between two
MCs can be understood by comparison to a billiards game. For one
moving rigid ball colliding with another still ball along the radial
direction, the response is straightforward in a vacuum: in the DC
case, two balls will strictly move along the same radial direction;
in the OC case, two balls will oppositely deflect along an angular
direction, accompanying their continuous radial movement. The
patterns of ball movement contribute to a further quantitative
understanding of complex collision between two MCs in the
interplanetary medium. The magnetic field lines, frozen in a low
$\beta$ plasma, could be considered as an elastic skeleton embedded
in the MC body. The innate magnetic elasticity can efficiently
buffer the compression as a result of the colliding of a following
MC2 against a leading MC1 \citep{Xiong2007a}. When every MC becomes
increasingly stiff, the compression reaches its asymptotic degree.
The compressibility effect should be included for the quantitative
investigation of deflection effect as a result of OC. Such a task
should use numerical MHD simulation, as we demonstrate in this
paper. The direction of main compression within the double MCs is
parallel/oblique to the radial direction for the DC/OC. For the DC
already discussed by \citet{Xiong2007a}, the compression strictly
persists along the heliospheric equator and the compressed magnetic
flux within the MC body almost points to the south. Therefore, the
DC case is very efficient to enhance the geoeffectiveness. The fast
MC2 body continuously strikes the slow MC1 tail, until the MC2 speed
is lower than the MC1 speed after momentum transfer. Such MC2 body
pushing prevents magnetic field lines in the MC1, previously
compressed by MC2-driven shock, from being restored, when the
MC2-driven shock completely passes through the MC1 body. For the OC,
the compression occurs along one side of each MC. For an example,
the MC2 body is faced with the MC1 body from its left side and the
ambient solar wind from its right side. Due to the MC1's blocking,
the MC2 suffers the compression from its left side. Such an angular
pressure imbalance leads to the MC2's right deflection.
Simultaneously, the MC1 deflects leftward for the same reason. The
opposite deflections, separating double MCs, greatly relieve the
intensity of the OC. Therefore, the angular freedom for each MC is
an extra factor in efficiently buffering the compression. This
angular deflection, absent for the DC case \citep{Xiong2007a}, is
explored for the OC case here. The IMF lines within the latitude
difference ($D\theta|_{t=0}=10^\circ$ in this case) of two MC
eruptions are first draped and then compressed between the MC1 tail
and MC2 head. At 22 hours, the left flank of the MC2-driven shock
enters the MC1 core (Figures \ref{case2-Vr}a and \ref{case2-Line}a).
Due to very low $\beta$ in the MC1 medium, the left flank of the
shock front in the MC1 body propagates much faster than the right
one in the ambient solar wind. The MC1 body is compressed by the
MC2-driven shock along its normal. The advance of the MC2-driven
shock accompanies a drastic jump of local speed in the MC1 medium.
The MC2 body obliquely chases the MC1 body and then grazes the MC1's
right boundary. During this process, the momentum is gradually
transferred from the following MC2 to the preceding MC1. The
location of the most violent interaction within the double MCs,
characterized by the greatest compression of local magnetic flux,
gradually shifts from the MC1's rear half (Figures \ref{case2-Vr}a
and \ref{case2-Line}a) to the MC1's anterior half (Figures
\ref{case2-Vr}b and \ref{case2-Line}b), and is finally within the
MC1-driven sheath (Figures \ref{case2-Vr}c and \ref{case2-Line}c).
The compound stream of double MCs reaches a relatively stable state
at 57 hours (Figures \ref{case2-Vr}c and \ref{case2-Line}c) when the
MC2-driven shock ultimately merges with the MC1-driven shock into a
stronger compound one.

The time sequence of synthetic measurement at L1 for case E$_2$ is
shown in Figure \ref{Satellite-2}. The speed $v_r$ monotonically
decreases from the MC1's head to the MC2's tail (Figure
\ref{Satellite-2}c). The magnetic elasticity of southward magnetic
flux takes a recovering effect against the previous compression,
as the MC2-driven shock continuously moves forward in the MC1
body. As the compression of the MC1's rear half is largely
relieved, the duration of geoeffective solar wind flow is
prolonged from 20 hours in case E$_1$ to 31 hours in case E$_2$.
Owing to the OC, the MC1 deflects northward, and the MC2 deviates
southward. Largely reduced is the total southward magnetic flux
passing through Lat. $=4.5^\circ$S. The opposite deflection of the
two MCs together with the above-mentioned mitigated compression
cause a significant increase of $Dst$ from $-234$ to $-121$ nT.

The evolution of various physical parameters for each MC in case
E$_2$ is shown in Figure \ref{geometry-evol}. The MC1 is
accelerated and the MC2 is decelerated, as seen in Figure
\ref{geometry-evol}a. The MC1 begins to deflect northward at 28
hours, 16 hours later than the MC2's southward deflection (Figure
\ref{geometry-evol}b). The deflections of both MC1 and MC2
gradually approach an asymptotic values
$\delta\theta_{mc1}=-3^\circ$ and $\delta\theta_{mc2}=6.3^\circ$,
respectively. Last, after being pushed aside, the MC1 and MC2
propagate along the latitudes of $\theta_{mc1}=3^\circ$N and
$\theta_{mc2}=16.3^\circ$S, respectively. Obviously, the MC2
undergoes a larger deflection than the MC1. Due to the deflection,
the distance $d$ between the two MC cores is highly increased in
contrast to an uncoupling case (Figure \ref{geometry-evol}c). For
an individual MC, the higher the MC speed is, the greater the
compression between the MC body and its front ambient solar wind,
and the smaller the MC's cross-section area $A_{mc}$. For an OC
case of double MCs, $A_{mc}$ depends on one more compression
factor, interaction between two MCs. With the increased speed, the
MC1 suffers larger compression from its front ambient solar wind.
The MC2's trailing pounding the MC1 compresses the MC1 body.
Therefore, the MC1 area $A_{mc1}$ is smaller than its
corresponding isolated case (Figure \ref{geometry-evol}d), which
is consistent with the DC case \citep{Xiong2007a}. However, the
MC2 area $A_{mc2}$ for the OC (Figure \ref{geometry-evol}e) is
quite contrary to that for the DC \citep[cf. Figure 5d
in][]{Xiong2007a}. The inconsistency is ascribed to the deflection
in the OC. For the DC, the compression of MC2 chiefly exists
between the MC1 tail and the MC2 head. The persistent blocking of
the MC1 body causes the MC2 area shrinkage. For the OC, the MC2
senses the compression from two aspects: (1) the front solar wind,
and (2) the sideward MC1 body. The MC2 slowdown tends to enhance
$A_{mc2}$, which indicates the mitigated compression between the
MC2 body and the solar wind. The blocking of the MC1 body at MC2's
left tends to reduce $A_{mc2}$. As the MC2 deflects sideward, the
former factor of the front solar wind contributes more to
$A_{mc2}$, and the latter factor of the sideward MC1 body
contributes less. The integration of both competing factors
determines the increase or decrease of $A_{mc2}$ in contrast to
its corresponding individual MC case. As for the current case
E$_2$, $A_{mc2}$ is increased.

\subsection{Latitudinal Distribution of Geoeffectiveness}
It has been substantiated from both observation data analyses
\citep[e.g.,][]{Burlaga1987,Wang2003a,Farrugia2006,Dasso2009} and
numerical simulations
\citep[e.g.,][]{Xiong2006a,Xiong2006b,Xiong2007a,Xiong2007b} that
multiple ICME interactions can significantly enhance the
geoeffectiveness at 1 AU. The near-equator latitudinal
distribution of $Dst$ index is plotted in Figure \ref{Dst-Lat}.
Since an individual MC would propagate radially through
interplanetary space, the MC core passage corresponds to the
strongest geomagnetic storm in a one-dip latitudinal distribution
of geoeffectiveness. The strongest geoeffectiveness is $-103$ nT
at $0^\circ$ for an isolated MC1 event (case $P_1$) and $-140$ nT
at $10^\circ$S for an isolated MC2 event (case $F_3$). The
coupling of two MCs obviously aggravates the geoeffectiveness. For
case E$_1$, the geoeffectiveness of the two MCs is overlapped, so
that the $\theta-Dst$ curve looks like a single dip with its
minimum $-262$ nT at $9^\circ$S. For case E$_2$, the initial delay
between the two MCs is short ($t_{mc2}=12.2$ hours), the double
MCs experience sufficient evolution, the accumulated deflection
angle becomes very pronounced, the latitudinal distance between
the two MCs becomes large, and the geoeffectiveness of the two MCs
is thus separated; hence the $\theta-Dst$ curve behaves like two
local dips with their local minima of $-200$ nT at $15^\circ$S and
$-145$ nT at $1.5^\circ$N. As the compound stream at 1 AU formed
by the two MCs' coupling evolves from case E$_1$ to E$_2$, its
geoeffectiveness is significantly diffused along the latitude with
the intensity largely reduced.

The interplanetary dynamics and resulting geoeffectiveness of the
double MCs is a complex system involving multiple independent
variables. The parametric studies of eruption-interval dependence
(EID), angular-difference dependence (ADD), and
collision-intensity dependence (CID) are further explored below to
continue our preliminary efforts for the cases E$_1$, E$_2$ of
double MCs, and the cases P$_1$, F$_3$ of a single MC. In sections
\ref{Sec:EID}, \ref{Sec:ADD}, and \ref{Sec:CID}, the
geoeffectiveness of double MCs is described by a scalar of minimum
$Dst$ along its latitudinal distribution.

\section{Eruption-Interval Dependence} \label{Sec:EID}
The results of eruption-interval dependence is elucidated in
Figure \ref{EID}. The transfer of momentum from the fast MC2 to
the slow MC1 leads to shortening of the Sun-Earth transient time
$TT_{mc1}$ and lengthening of $TT_{mc2}$ (Figure \ref{EID}a). As
the initial eruption delay ($t_{mc2}-t_{mc1}$) is shortened, the
deflection of each MC exhibits an asymptotic behavior (Figure
\ref{EID}b). When $t_{mc2}$ is reduced from 33.1 to 22.1 hours,
and then to 10.2 hours, the MC1 deflection angle
$|\delta\theta_{mc1}|$ is increased from $0.2^\circ$ to
$0.7^\circ$, and then to $3.3^\circ$; the MC2 deflection angle
$|\delta\theta_{mc2}|$ is enhanced from $1.2^\circ$ to
$2.7^\circ$, and then to $6.9^\circ$; the total deflection angle
$\Delta\theta$
($\Delta\theta=|\delta\theta_{mc1}|+|\delta\theta_{mc2}|$) is
changed from $1.4^\circ$ to $3.4^\circ$, and then to $10.2^\circ$.
These deflection ratios of
$|\delta\theta_{mc1}|:|\delta\theta_{mc2}|$ at $t_{mc2}=$ 33.1,
22.1, and 10.2 hours correspond to 0.17, 0.26, and 0.48,
respectively. Obviously, the MC2 occupies a much bigger share of
the total deflection angle $\Delta \theta$. This latitudinal
deflection is manifested in the distance $d$ between the cores of
MC1 and MC2 at 1 AU (Figure \ref{EID}d). At the beginning, the
magnetic field lines in the MC1 rear half are too vulnerable to
resist the MC2's pounding, so the temporarily enhanced compression
leads to $d$ decrease. As interpreted in section \ref{Sec:Case2},
when the most intensely interacting region in the double MCs is
shifted from the MC1's rear half, the magnetic elasticity, being
passively quenched at the earlier time by the MC2-driven shock,
begins to actively bounce to push the two MCs apart. Then the $d$
is steadily increased. Hence the minimum $d$ of 63 $R_s$ exists in
an intermediate platform of $t_{mc2}=22 \sim 28$ hours. Between
this zone of $t_{mc2}=22 \sim 28$ hours, the double MCs suffer the
strongest compression, the MC1's cross section area $A_{mc1}$ is
compressed to its minimum $2.4 \times 10^3$ $R_s^2$ at
$t_{mc2}=22$ hours (Figure \ref{EID}e), and $Dst$ reaches a
minimum of $-270$ nT at $t_{mc2}=28$ hours (Figure \ref{EID}f).
When $t_{mc2} \le 22$ hours, the magnetic elasticity restoration
and angular deflection lead to the increasingly weak compression
between two clouds and a consequent increase in $A_{mc1}$. In
addition, according to the reasons given in section
\ref{Sec:Case2} for the $A_{mc2}$ variance, the predominance of
the MC2's momentum loss over the MC1's blocking is responsible for
the monotonic increase of $A_{mc2}$ between $Dt= -32 \sim -10$
hours. These behaviors of the OC case are essentially different
from those of the DC case \citep{Xiong2007a}. For the DC case
without the deflection effect, the persistent following of the MC2
body at the MC1 tail can inhibit the MC1 body from re-expanding,
so the $Dst$ can be roughly maintained at a constant, given that
the initial delay $t_{mc2}$ is smaller than a certain threshold
\cite[cf. Figure 9 in][]{Xiong2007a}. The perpetual balance
between the external compression and innate elasticity for the DC
case is out of equilibrium for the OC case under a new
circumstance of angular deflection. Moreover, the external
compression could be largely offset by the deflection. Therefore,
the OC case is generally weaker for geoeffectiveness than its
corresponding DC case, and the strongest geoeffectiveness for the
OC case can only be achieved at a certain initial delay between
two MC eruptions.

\section{Angular-Difference Dependence} \label{Sec:ADD}
The angular-difference dependence is shown in Figure \ref{ADD}.
The initial latitudinal difference of
$D\theta|_{t=0}=\theta_{mc2}|_{t=0} - \theta_{mc1}|_{t=0}$ for
solar output decides the oblique degree of interplanetary
collision between two MCs. Corresponding to the nonexistence of
deflection, $D\theta|_{t=0}=0^\circ$ is due to the symmetrical
condition, which was thoroughly addressed by \citet{Xiong2007a}.
When $D\theta|_{t=0}$ is too large, the OC effect will be
significantly mitigated, and the consequent deflection will be
obviously weak. An appropriate $D\theta|_{t=0}$ corresponds to the
maximum deflection of OC cases of double MCs, very similar to the
known conclusion for the OC case of ``a shock overtaking an MC"
\citep{Xiong2006b}. At $D\theta|_{t=0}=15^\circ$, the total
deflection angle $\Delta\theta$ reaches its maximum $12.2^\circ$
with $\delta\theta_{mc1}=-3.8^\circ$ and
$\delta\theta_{mc2}=8.4^\circ$. The $|\delta\theta_{mc2}|$ is
generally larger than the $|\delta\theta_{mc1}|$, but it does not
match the case of $D\theta|_{t=0} > 40^\circ$. When
$D\theta|_{t=0}
> 40^\circ$, the two MCs are so widely separated that the
interaction is virtually ascribed to the coupling of the MC1 body
and the MC2-driven shock. Such indirect interaction between the
two MC bodies to transfer momentum clarifies $|\delta\theta_{mc1}|
> |\delta\theta_{mc2}|$ for $D\theta|_{t=0}=40^\circ \sim
50^\circ$ (Figure \ref{ADD}b) and the $A_{mc2}$ decrease for
$D\theta|_{t=0} = 20^\circ \sim 50^\circ$ (Figure \ref{ADD}e).
With respect to the $A_{mc2}$ variance for $D\theta|_{t=0} \le
20^\circ$, two competing factors of MC2 momentum loss and MC1 body
blocking take effect, as previously interpreted in section
\ref{Sec:Case2}. The dominance of the MC2 momentum loss accounts
for the increase between $D\theta|_{t=0}=10^\circ \sim 20^\circ$;
that of the MC1 body blocking elucidates the decrease between
$D\theta|_{t=0}=0^\circ \sim 10^\circ$. The closer the two MCs are
in the near-Sun position, the smaller the distance $d$ at 1 AU
(Figure \ref{ADD}d). As $D\theta|_{t=0}$ decreases, the $Dst$,
being steadily reduced with a steeper slope, changes from $-140$
nT at $D\theta|_{t=0}=50^\circ$ to $-230$ nT at
$D\theta|_{t=0}=0^\circ$. The less the deflection effect is, the
more compact the multiple interplanetary geoeffective triggers and
the more violent the ensuing geomagnetic storm.

\section{Collision-Intensity Dependence} \label{Sec:CID}
Figure \ref{CID} displays the collision-intensity dependence. The
variance of $v_{mc2}$ corresponds to a different individual MC2
event. As $v_{mc2}$ increases, both $TT_{mc1}$ and $TT_{mc2}$
decrease. However, the decreased $TT_{mc2}$ in the case of double
MCs is still larger than its corresponding individual MC case
(Figure \ref{CID}a). The influence of the OC intensity can be
described by the $v_{mc2}$ in some senses. As an asymptotic
response to the $v_{mc2}$ increase from 450 to 1200 km/s, the
geoeffective $Dst$ decreases from $-145$ to $-255$ nT (Figure
\ref{CID}f), and the total deflection angle $\Delta\theta$
increases from $7.5^\circ$ to $12.8^\circ$ (Figure \ref{CID}c).
The contribution of $\Delta\theta$ almost stems from the MC1
deflection $\delta_{mc1}$, since the MC2 deflection angle
$\delta_{mc2}$ is nearly constant at $6^\circ$ (Figure
\ref{CID}b). The deflection ratio between the MC1 and MC2
$|\delta\theta_{mc1}|:|\delta\theta_{mc2}|$ is 0.3 at
$v_{mc2}=450$ km/s, 0.7 at $v_{mc2}=800$ km/s, and 1 at
$v_{mc2}=1200$ km/s. Therefore, the cause of intensity aggravation
of two MCs' colliding is mainly manifested in the response of the
preceding MC1 body. In addition, for the case of
$D\theta|_{t=0}=10^\circ$ in group ADD, the MC2's cross-section
area $A_{mc2}$ of a coupled case is larger than that of an isolate
case (Figure \ref{ADD}e). Since $D\theta|_{t=0}$ equals $10^\circ$
in group CID, the $A_{mc2}$ behavior is similar for the reason
explained in section \ref{Sec:ADD}. Furthermore, the more intense
the OC is between two MCs, the more violent the compression in the
double MCs, and the stronger the accumulated innate magnetic
elasticity against the external compression. So the deflection
angle $\delta\theta$ (Figure \ref{CID}b), the distance between the
two MC cores $d$ (Figure \ref{CID}d), and the geoeffective $Dst$
(Figure \ref{CID}f) all exhibit an asymptotic behavior.

\section{Conclusions and Summary}\label{Sec:Conclusion}
The dynamics and geoeffectiveness of interplanetary compound
structures such as the complex ejecta \citep{Xiong2005}, MC-shock
\citep{Xiong2006a,Xiong2006b}, and MC-MC \citep{Xiong2007a} have
been comprehensively investigated during the recent years with our
2.5-D numerical model within an ideal MHD framework. As a logically
direct continuation to the DC mode between a preceding MC1 and a
following MC2 \citep{Xiong2007a}, the OC mode is further explored
here to highlight a deflection effect from the parametric studies of
eruption-interval dependence, angular-difference dependence, and
collision-intensity dependence. The deflection angle for an MC1-MC2
OC in this paper is obviously greater than that for an MC-shock OC
addressed by \citet{Xiong2006b}, as the MC1-MC2 coupling involves a
comprehensive interaction among the MC1-driven shock, the MC1 body,
the MC2-driven shock, and the MC2 body.

An interplanetary compound stream is formed as a result of
interaction between two MCs in the interplanetary space. The
direction of main compression within the double MCs is
parallel/oblique to the radial direction for the DC/OC. The OC
leads to first compress each MC on one side, then push the MC to
the other side as a result of angular pressure imbalance. Such a
deflection effect for the OC case is essentially absent for the DC
case. The deflection angles of MC1 ($|\delta\theta_1|$) and MC2
($|\delta\theta_2|$) asymptotically approach their corresponding
limits, when the two MC-driven shocks are merged into a stronger
compound shock. During this process, the geoeffectiveness of
double MCs is significantly diffused along the latitudinal
distribution, with the intensity largely reduced. An appropriate
angular difference between the initial eruptions of two MCs leads
to the maximum deflection of $|\delta\theta_1|$ and
$|\delta\theta_2|$. A continuous increase of OC intensity can
synchronously enhance $|\delta\theta_1|$ and $|\delta\theta_2|$,
although its effect becomes less and less obvious. The response of
$|\delta\theta_1|$ is far more sensitive than that of
$|\delta\theta_2|$. The $|\delta\theta_1|$ is generally far less
than the $|\delta\theta_2|$, and the unusual case of
$|\delta\theta_1| \simeq |\delta\theta_2|$ only occurs for the
extremely intense OC. The opposite deflection between two MCs,
together with the inherent magnetic elasticity of each MC, could
efficiently buffer the external compression for the interplanetary
OC.

The axial variance of an MC is ignored in our model for
simplification, so that the geometry of an MC is reduced to be
2.5-D. In reality, both feet of an interplanetary MC is still
connected to the solar surface, as substantiated from the evidence
of bi-directional electron fluxes along an MC's axis
\citep{Larson1997}. However, for the local analyses of a cross
section of an MC, a locally cylindric flux-rope has widely been used
to approximate the globally curved one, such as the data inversion
from the near-Earth in-situ observations \citep{Burlaga1981}, the
kinematic model of an MC propagation \citep{Owens2006}, and the
numerical simulation of magnetic-flux-rope dynamics
\citep{Schmidt2004,Xiong2006a}. Hence, our 2.5-D model can well
reflect some dynamic characteristics of 3-D MCs to some extent.

An assimilatively integrated study of observation data analyses and
numerical simulations is crucial and effective for an in-depth and
overall understanding of the Sun-Earth system. As a CME is 3-D by
nature, a 2.5-D model has serious limitations in space weather
predicting, and a full 3-D numerical model is indispensable to
describe realistic observation events. On the one hand, data-driven
3-D models can be tested and improved by using observation data; on
the other hand, observations can be better interpreted by using
global 3-D models. For instance, demonstrating a good match between
synthetic and real {\it STEREO/SECCHI} images, \citet{Lugaz2009}
quantitatively analyzed and well explained the January 24-25 CME
event by a data-driven 3-D numerical MHD model. Hence, two MCs'
interacting in the 2.5-D model in this paper is meaningfully
generalized to a 3-D geometry. Such model generalization and then
detailed comparison with realistic events are out of contents in
this paper and will be addressed in our near future.

In closing, the interaction among multiple CMEs/ICMEs can be a
cause of angular deflection during the CME/ICME propagation. Such
angular displacement, being nonlinear and irreversible, results in
the significant responses of interplanetary dynamics and ensuing
geoeffectiveness. Therefore, when successive CMEs from the solar
corona are likely to collide with each other obliquely in the
interplanetary space, the factor of potential deflection due to
the OC should be considered for the geoeffectiveness prediction at
1 AU, as well as the correlation between the near-Sun and the
near-Earth observations.

\begin{acknowledgments}
We are highly grateful to Drs. Amitava Bhattacharjee and Clia
Goodwin for their sincere and beneficial help in polishing the
language of our manuscript. This work was supported by the
National Natural Science Foundation of China (40774077), the
National Key Basic Research Special Foundation of China
(2006CB806304), the China Postdoctoral Science Foundation
(20070420725), and the K. C. Wong Education Foundation of Hong
Kong.
\end{acknowledgments}

\bibliography{Xiong}

%\clearpage
%\setfigurenum{1}
\newpage
\begin{table}
\vspace*{2cm} %
\caption{Assortment of simulation cases of an individual
MC}\label{Tab1}
%\begin{small}
\begin{tabular}{|c|c|c|c|}
\hline
Group & Case & $v_{mc}$ & Comment\\[0pt]
&& ($10^2$ km/s) &\\[0pt]
\hline
IPM & P$_1$  & 4  & Individual-Preceding MC \\[0pt]
&  &  & ($H_{mc}=1$) \\[0pt]
\hline
IFM & F$_1$, F$_2$, F$_3$, & 4.5, 5, 6, & Individual-Following MC \\[0pt]
& F$_4$, F$_5$, F$_6$, & 7, 8, 9, & ($H_{mc}=-1$) \\[0pt]
& F$_7$, F$_8$, F$_9$ & 10, 11, 12 &\\[0pt]
\hline  %----------------------------------------------------- Individual MC
\end{tabular}
%\end{small}
\end{table}

\newpage
\begin{table}[rotate=90] %\vspace*{-3.2cm}
\vspace*{2cm}%
\caption{Assortment of simulation cases of double MCs. Note that
$v_{mc1} = 400$ km/s, $\theta_{mc1}=0^\circ$, $t_{mc1}= 0$ hour,
$H_{mc1}=1$, $H_{mc2}=-1$ for all thirty-four cases.}\label{Tab2}
%\begin{small}
\begin{tabular}{|c|c|c|c|c|c|}
\hline
Group & Case & $v_{mc2}$ & $\theta_{mc2}$ & $t_{mc2}$ & Comment\\[0pt]
&&($10^2$ km/s) & (degree) & (hour) &\\[0pt]
\hline  %-----------------------------------------------------
EID & E$_1$, E$_2$, E$_3$, E$_4$, & 6 & 10 & 30.1, 12.2,
44.1, 42.1, & Eruption- \\[0pt]
& E$_5$, E$_6$, E$_7$, E$_8$, &&& 40.2, 37.2, 35.1,
33.1, & Interval \\[0pt]
& E$_9$, E$_{10}$, E$_{11}$, E$_{12}$, &&& 31.5, 28.2,
25.1, 22.1, & Dependence \\[0pt]
& E$_{13}$, E$_{14}$, E$_{15}$, E$_{16}$ &&& 20.1, 17.1,
15.1, 10.2 & \\[0pt]
\hline  %--------------------------------------------------------- Eruption Interval Dependence (EID)
ADD & A$_1$, A$_2$, A$_3$, E$_2$,, & 6 & 0, 3, 5,
10, & 12.2 & Angular- \\[0pt]
& A$_4$, A$_5$, A$_6$, A$_7$,  && 15, 20, 25,
30, && Difference \\[0pt]
& A$_8$, A$_9$, A$_{10}$ && 40, 45, 50 && Dependence \\[0pt]
\hline  %--------------------------------------------------------- Collision Intensity Dependence (CID)
CID & C$_1$, C$_2$, E$_2$, C$_3$, & 4.5, 5, 6, 7 & 10 & 12.2 & Collision- \\[0pt]
& C$_4$, C$_5$, C$_6$, C$_7$,  & 8, 9, 10, 11 & && Intensity \\[0pt]
& C$_8$ & 12 & && Dependence \\[0pt]
\hline
\end{tabular}
%\end{small}
\end{table}

%\iffalse
% double MCs: Case 1  ----------------------------------------------------------------------------------------------------------------------
\clearpage
\newpage

\setcounter{figure}{0}
\newpage
\begin{figure}
\noindent \caption{Evolution of an MC2 overtaking an MC1 for Case
E$_1$, with (a)-(c) magnetic field magnitude $B$. The white solid
line denotes the MC boundary. The white radial dashed line is along
the latitude of $4.5^\circ$. Only the part of domain is adaptively
plotted to highlight the double MCs.}
\end{figure}

\begin{figure*}
\noindent
  \includegraphics[width=0.93\textheight,angle=90]{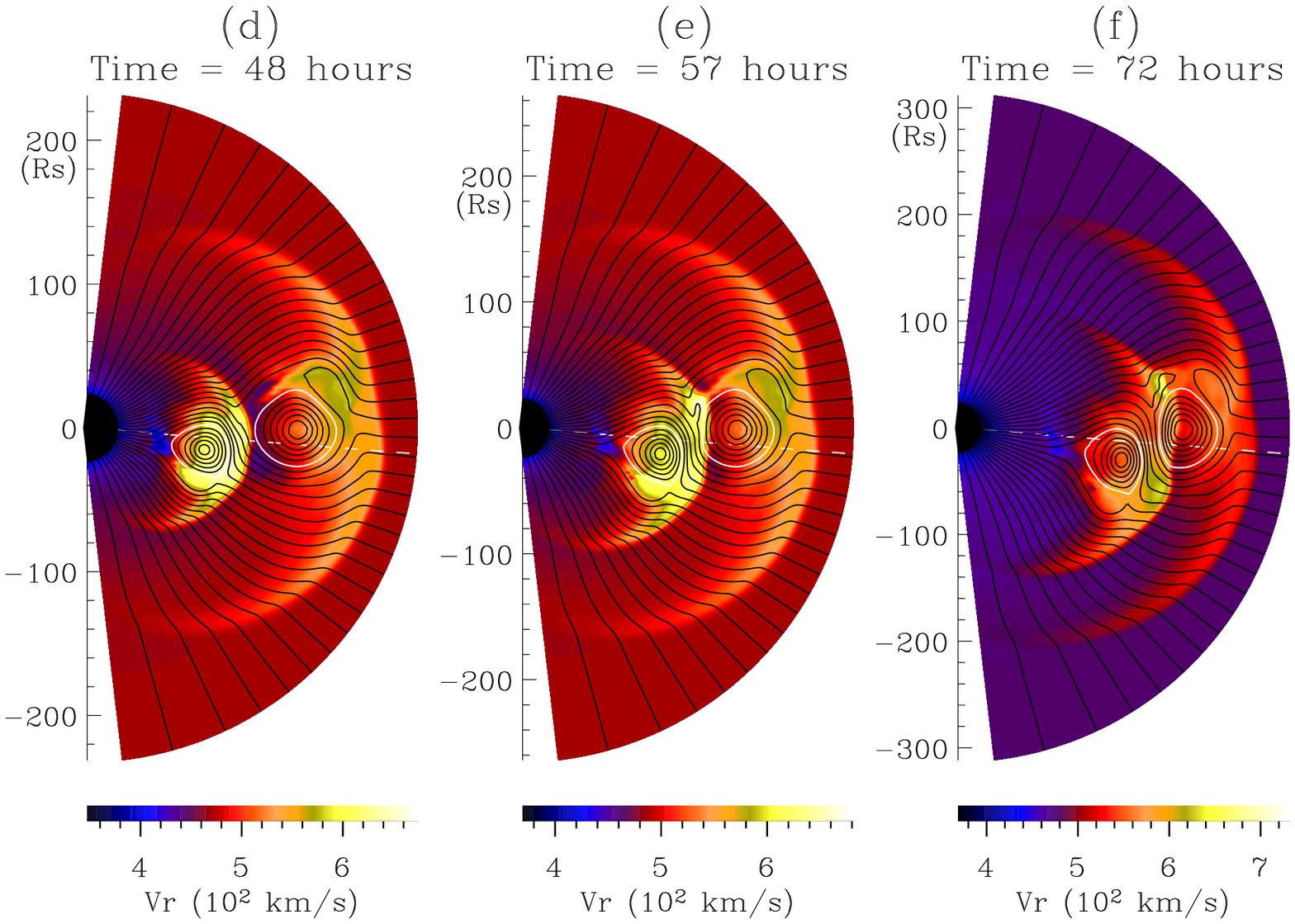}
\caption{} \label{case1-Vr}
\end{figure*}

\setcounter{figure}{1}
\newpage
\begin{figure}
\noindent \caption{Evolution of an MC2 overtaking an MC1 for Case
E$_1$, with (d)-(f) radial flow speed $v_r$.}
\end{figure}

\begin{figure*}
\noindent
  \includegraphics[width=0.93\textheight,angle=90]{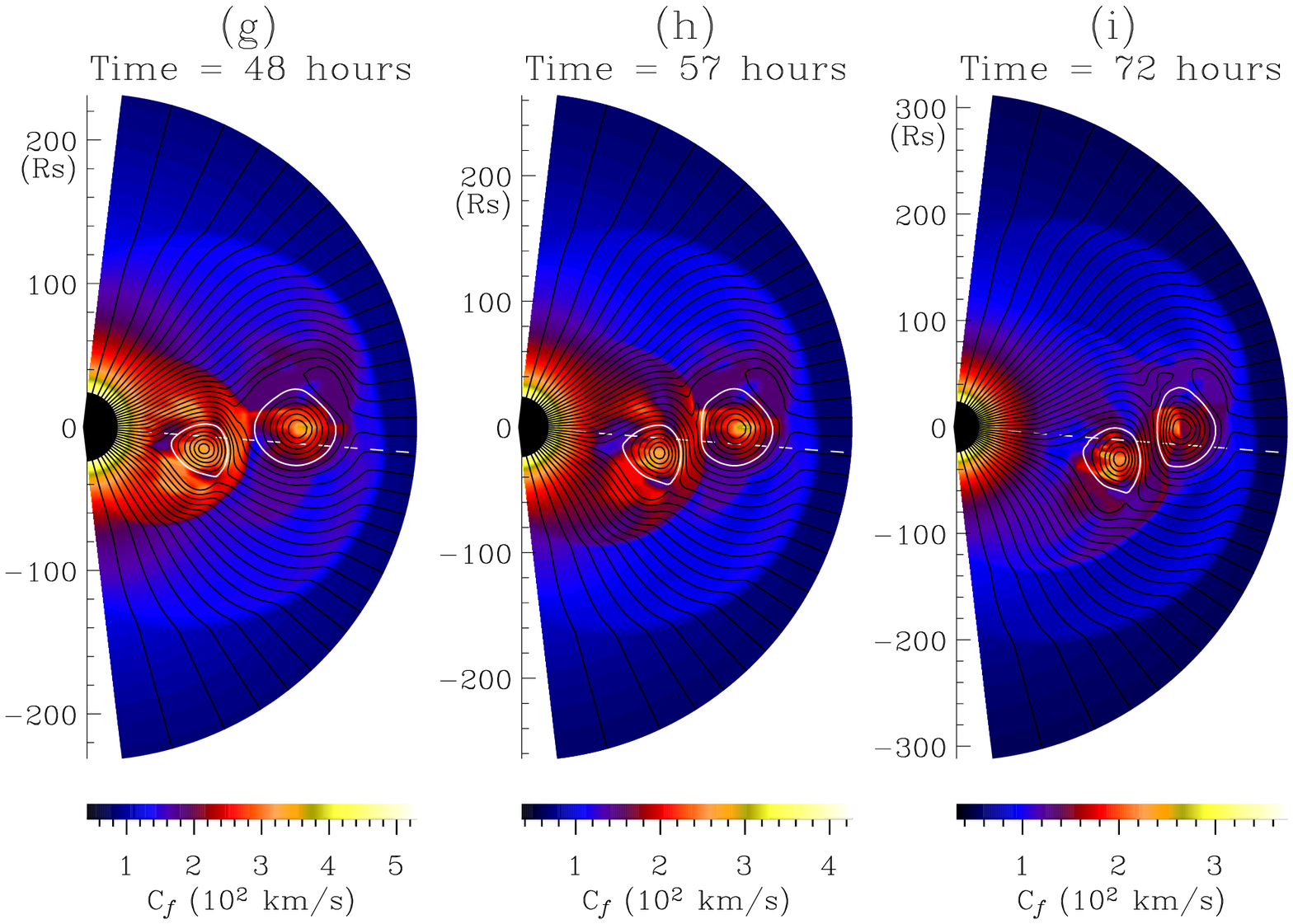}
\caption{} \label{case1-Cf}
\end{figure*}

\setcounter{figure}{2}
\newpage
\begin{figure}
\noindent \caption{Evolution of an MC2 overtaking an MC1 for Case
E$_1$, with (g)-(i) radial characteristic speed of fast mode $c_f$.}
\end{figure}

\begin{figure*}
\noindent
  \includegraphics[width=0.99\textheight,angle=90]{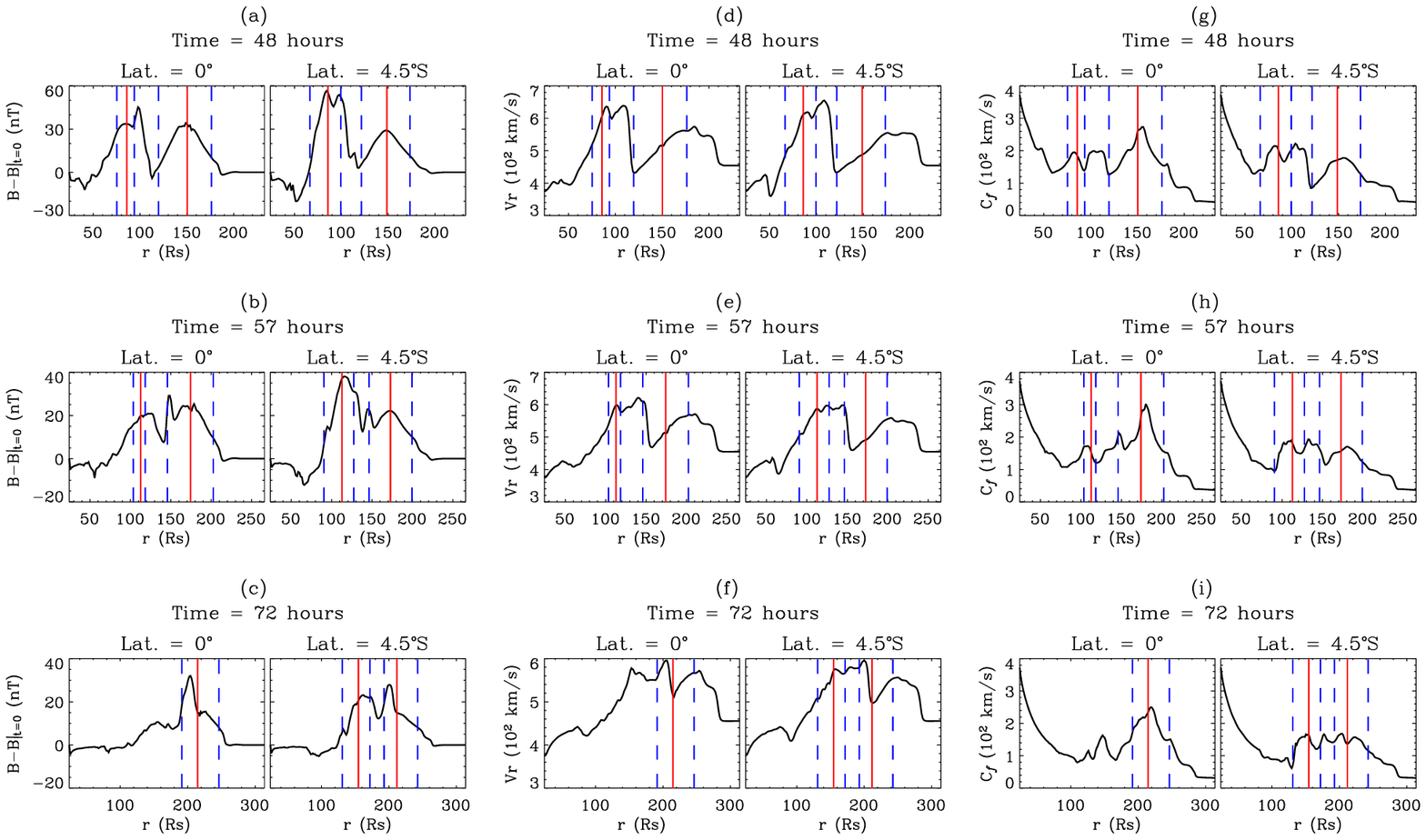}
\caption{} \label{case1-Line}
\end{figure*}

\setcounter{figure}{3}
\newpage
\begin{figure}
\noindent \caption{Two radial profiles along Lat.$=0^\circ$ and
$4.5^\circ$S for Case E$_1$. Note that radial profile of $B$ is
plotted by subtracting the initial ambient value $B|_{t=0}$. The
solid and dashed lines at each profile denote the MC core and
boundary, respectively.}
\end{figure}

\newpage
\begin{figure*}
\noindent
  \includegraphics[width=20pc]{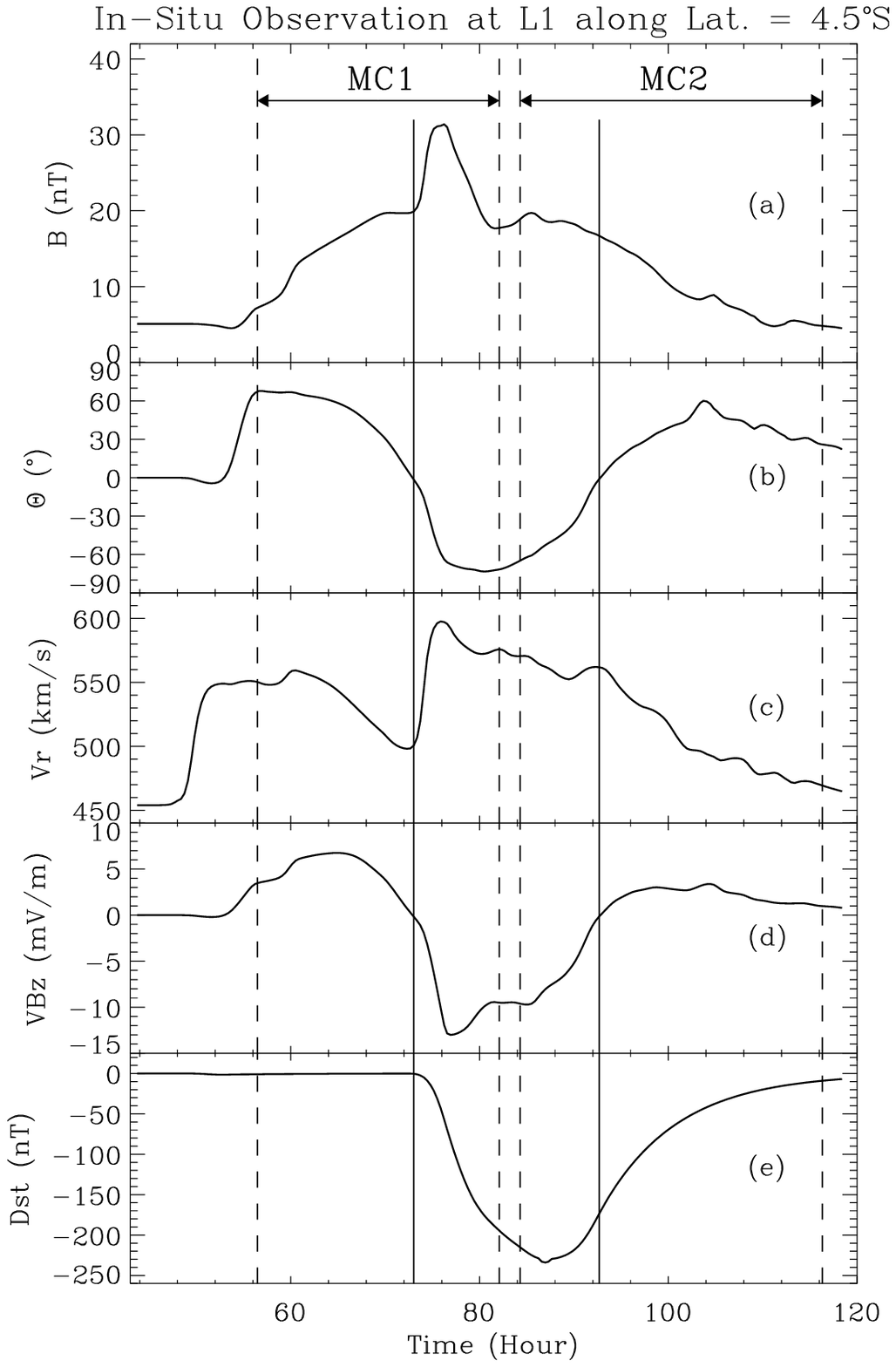}
\caption{In-situ synthetic observations along Lat. = $4.5^\circ$S
for Case E$_1$. Stacked from top to bottom are the (a) magnetic
field magnitude $B$, (b) elevation of magnetic field $\Theta$, (c)
radial flow speed $v_r$, (d) dawn-dusk electric field $VB_z$, and
(e) $Dst$ index. The solid and dashed delimiting lines denote the MC
core and boundary, respectively.} \label{Satellite-1}
\end{figure*}

% double MCs: Case 2  ----------------------------------------------------------------------------------------------------------------------
\newpage
\begin{figure*}
\noindent
  \includegraphics[width=0.93\textheight,angle=90]{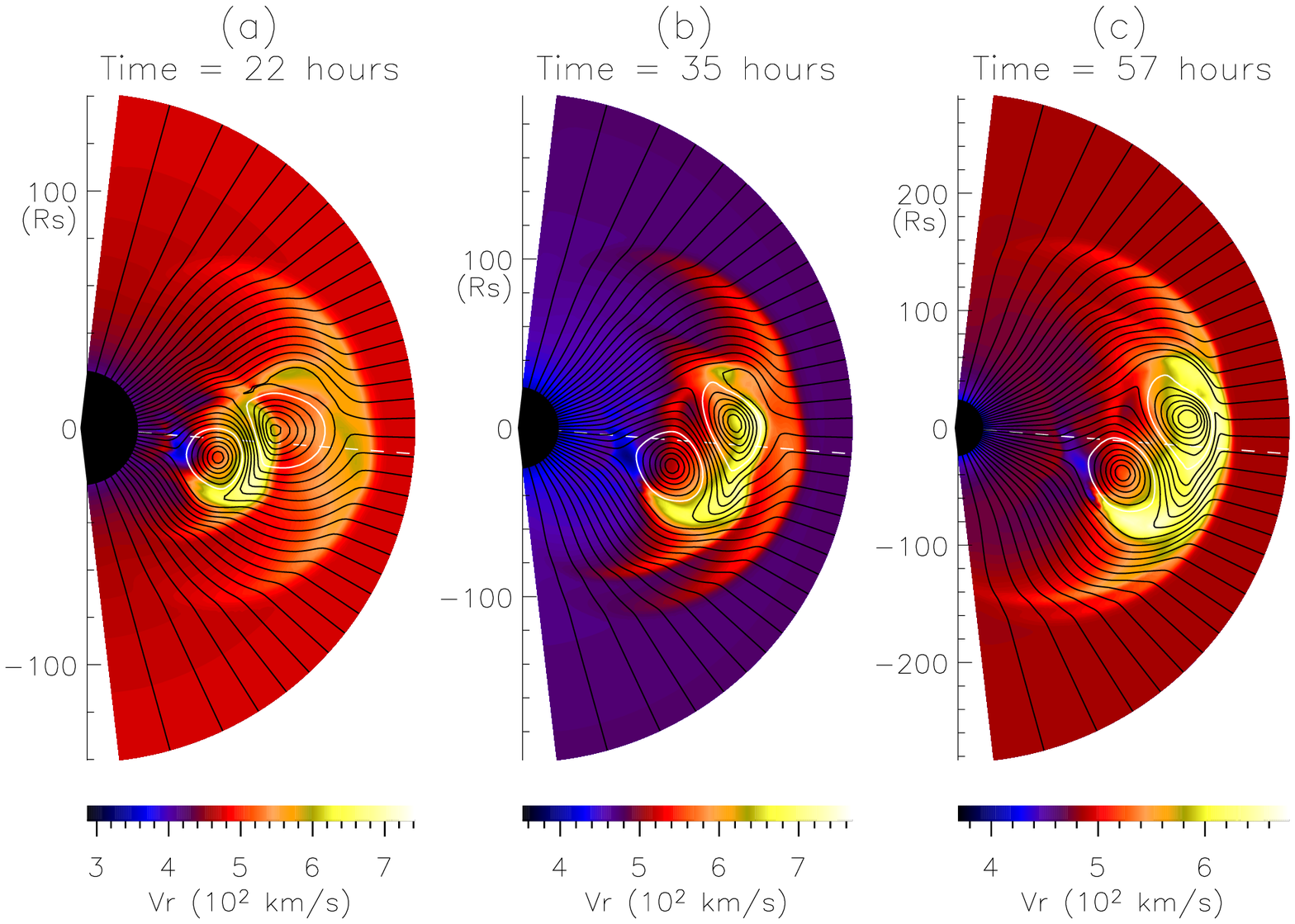}
\caption{} \label{case2-Vr}
\end{figure*}

\setcounter{figure}{5}
\newpage
\begin{figure}
\noindent \caption{Evolution of an MC2 overtaking an MC1 for Case
E$_2$, with radial flow speed $v_r$.}
\end{figure}

\begin{figure}
  \includegraphics[width=0.73\textwidth]{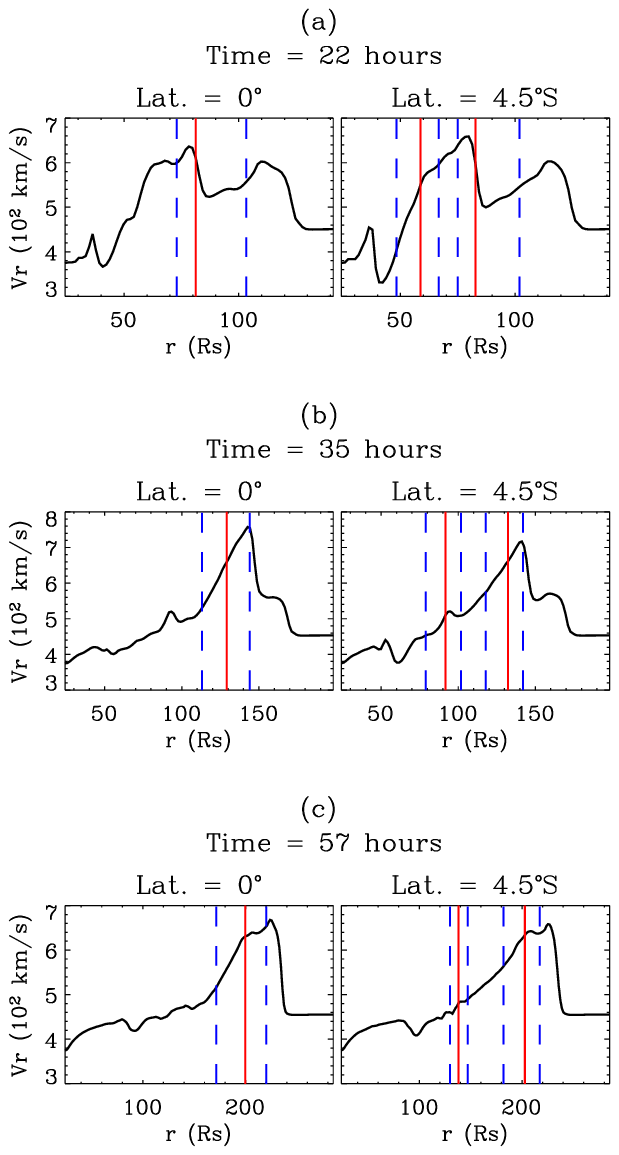}
\caption{Two radial profiles along Lat.$=0^\circ$ and $4.5^\circ$S
for Case E$_2$.} \label{case2-Line}
\end{figure}

\newpage
\begin{figure}
\noindent
  \includegraphics[width=20pc]{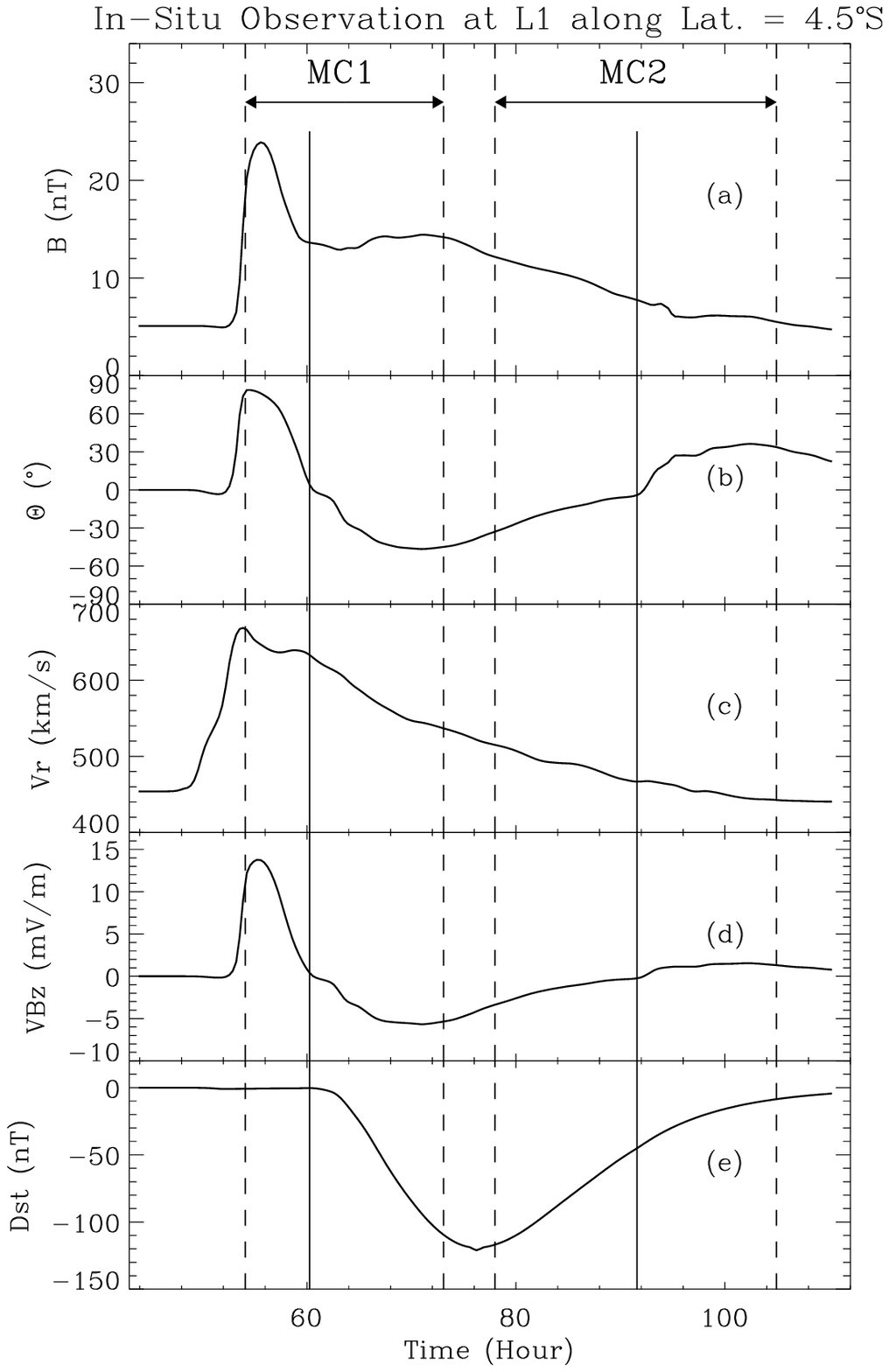}
\caption{In-situ synthetic observations along Lat. = $4.5^\circ$S
for Case E$_2$.} \label{Satellite-2}
\end{figure}

% Comparison among Cases E1, E2 ----------------------------------------------------------------------------------------
\newpage
\begin{figure}
  \includegraphics[width=20pc]{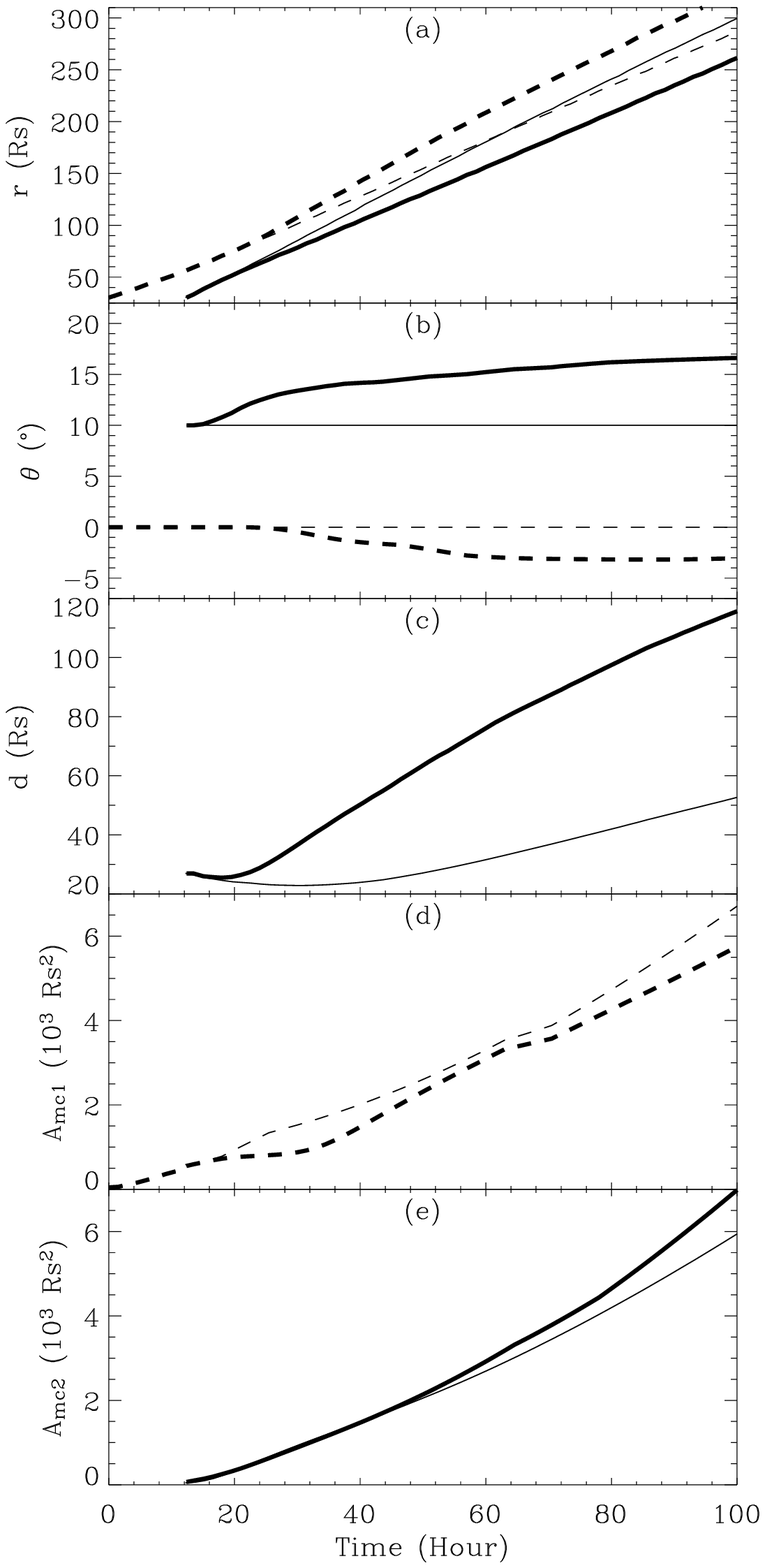}
\caption{Time dependence of MC parameters: (a) radial distance of MC
core $r$, (b) latitude of MC core $\theta$, (c) distance between
both MC cores $d$, (d) MC1 cross section area $A_{mc1}$, (e) MC2
cross section area $A_{mc2}$. In panels (a,b,d,e) the thick dashed
and solid lines denote the preceding MC1 and following MC2 in the
Case E$_2$, superimposed with the thin lines for the corresponding
individual MC cases for contrast. In panel (c) the thick and thin
lines represent the coupling and non-coupling conditions between two
MCs.} \label{geometry-evol}
\end{figure}

\newpage
\begin{figure}
  \includegraphics[width=0.85\textwidth]{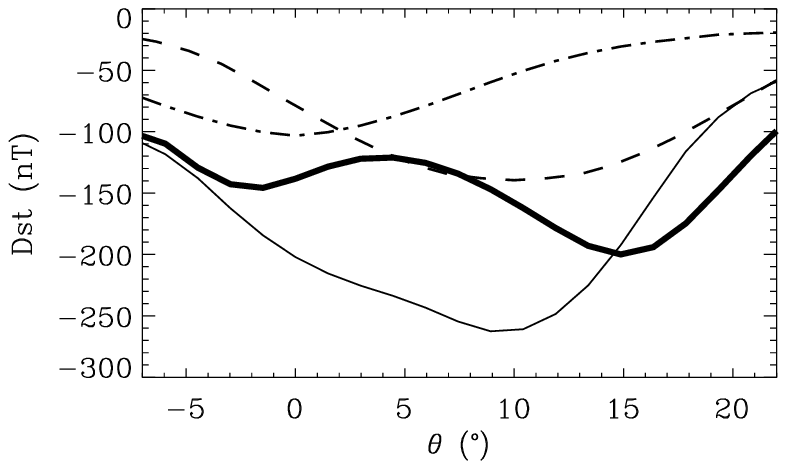}
\caption{Comparison of latitudinal distribution of $Dst$ index among
the compound-stream Cases E$_1$ (thin solid), E$_2$ (thick solid),
and corresponding individual-MC cases P$_1$ (dash-dotted), F$_3$
(dashed).} \label{Dst-Lat}
\end{figure}

% Synoptic Analysis for Multi-Cases  ------------------------------------------------------------
\newpage
\begin{figure}[htbp]
  \includegraphics[height=.99\textwidth,width=.95\textheight,keepaspectratio]{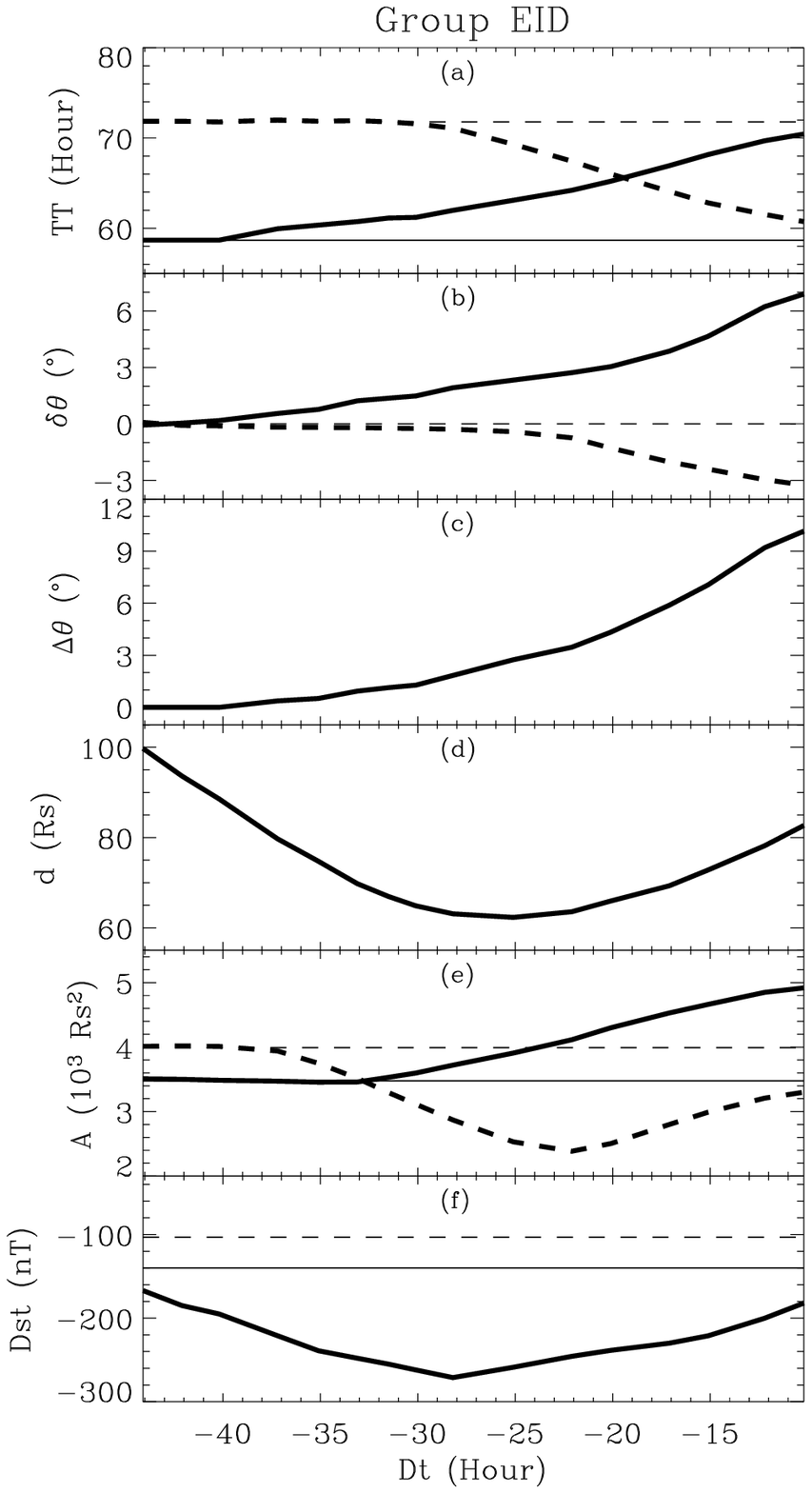}
\caption{Dependence of the compound-stream parameters at 1 AU on the
MC1-MC2 eruption delay $Dt$ ($Dt = t_{mc1} - t_{mc2}$) in group EID:
the (a) Sun-Earth transient time $TT$, (b) deflection angle of each
MC $\delta \theta$, (c) total deflection angle of double MCs $\Delta
\theta$ ($\Delta \theta= |\delta \theta_{mc1}| + |\delta
\theta_{mc2}|$), (d) distance between the two MC cores $d$ when the
MC1 core reaches 1 AU, (e) cross section area of each MC $A$, (f)
$Dst$ index. The thick dashed/solid lines in panels (a,b,e,f) refer
to the occasion of MC1/MC2 core reaching 1 AU. The thin dashed and
solid lines in panels (a,b,e,f) denote the isolated MC1 and MC2
events for comparison.} \label{EID}
\end{figure}

\newpage
\begin{figure}[htbp]
  \includegraphics[height=.99\textwidth,width=.95\textheight,keepaspectratio]{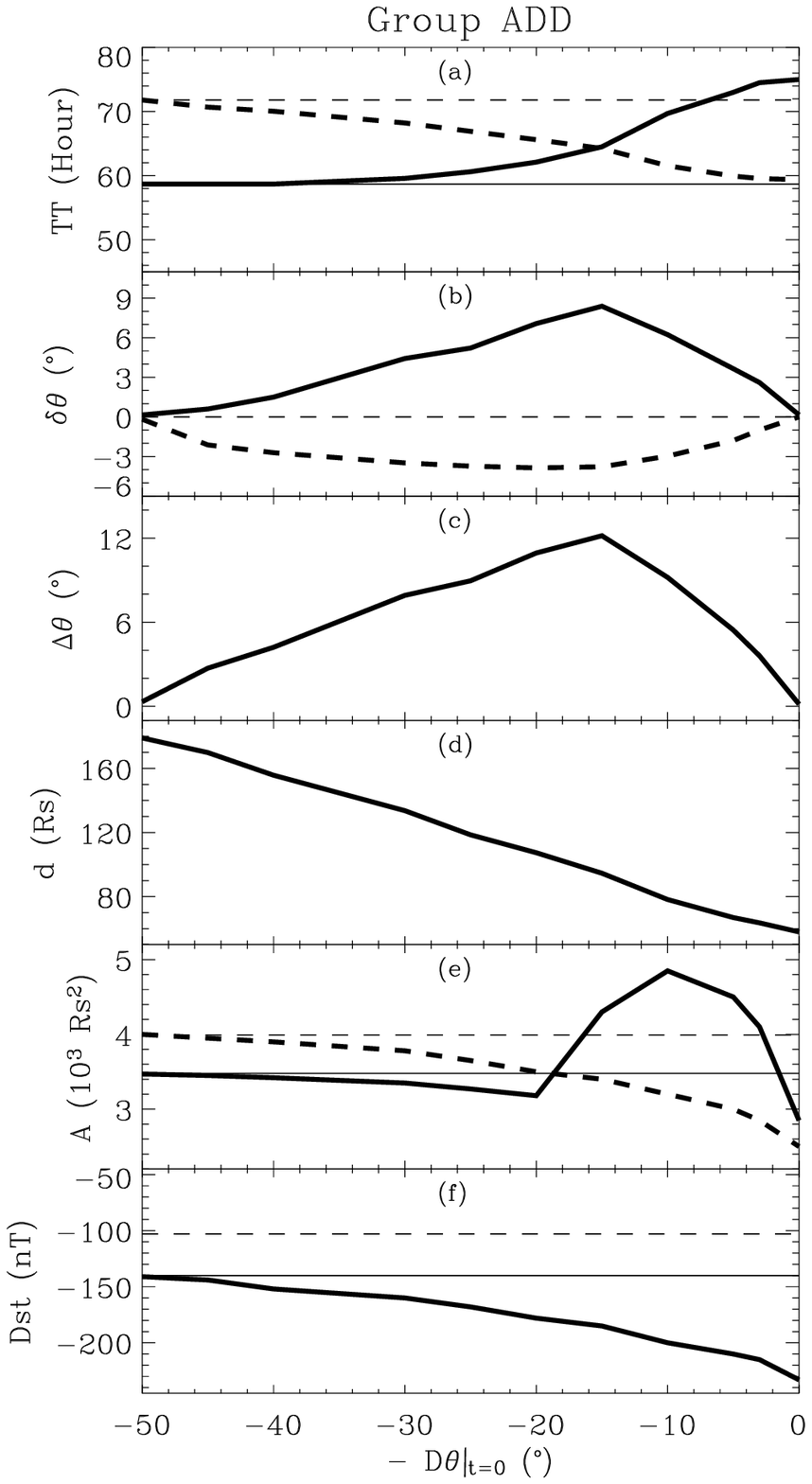}
\caption{Dependence of the compound-stream parameters at 1 AU on the
angular difference $-D \theta |_{t=0}$ of the two MC eruptions in
group ADD. Here $-D \theta |_{t=0} = -1 \cdot D \theta |_{t=0} =
\theta_{mc1}|_{t=0} - \theta_{mc2}|_{t=0}$.} \label{ADD}
\end{figure}

\newpage
\begin{figure}[htbp]
  \includegraphics[height=.99\textwidth,width=.95\textheight,keepaspectratio]{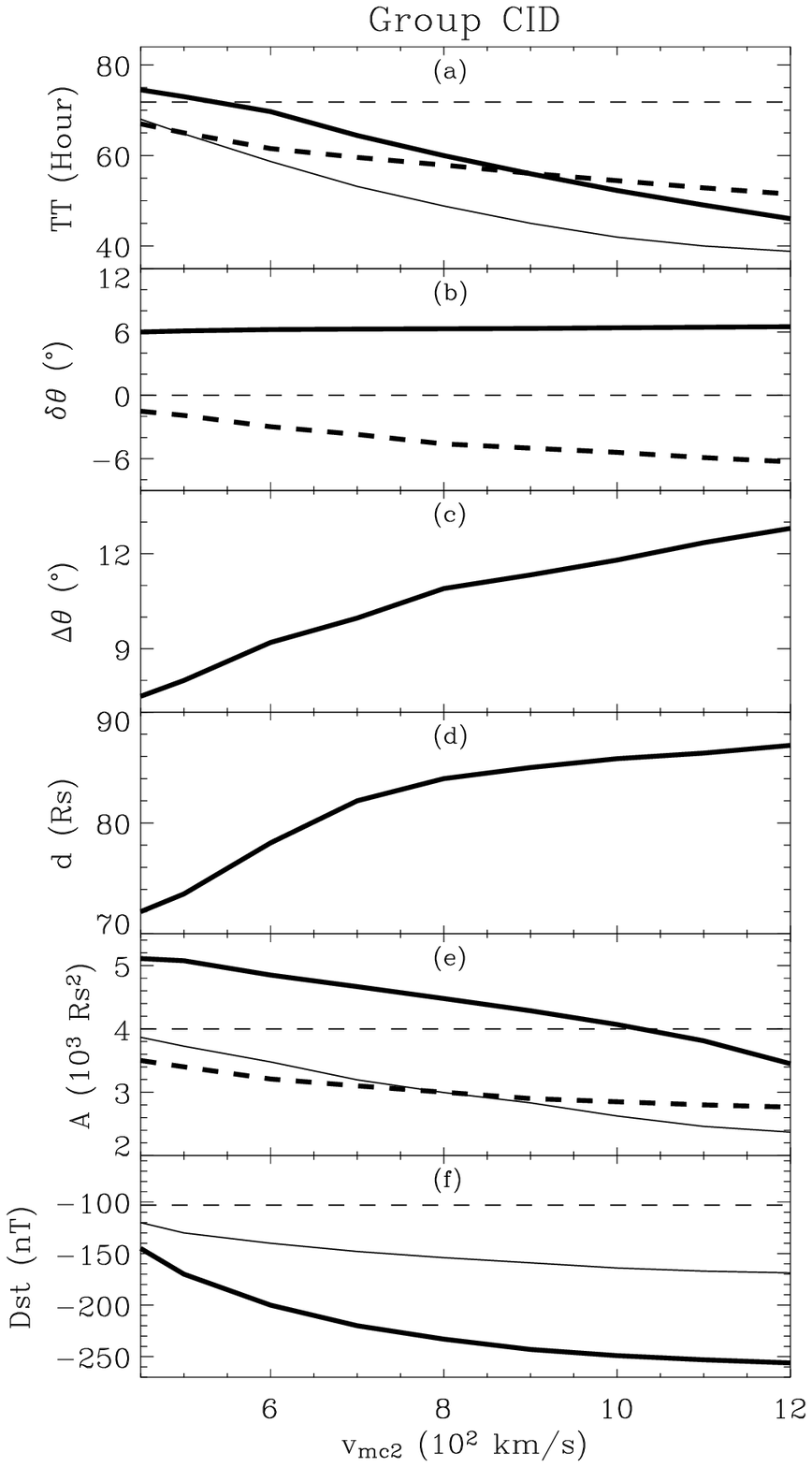}
\caption{Dependence of the compound-stream parameters at 1 AU on the
MC2 eruption speed $v_{mc2}$ in group CID.} \label{CID}
\end{figure}
%\fi

\clearpage

\end{article}
\end{document}